\documentclass[preprint, aps,amsmath,byrevtex,prd, titlepage, superscriptaddress,
nofootinbib]{revtex4-1}

\usepackage{dcolumn}
\usepackage{amsmath}
\usepackage{amssymb}
\usepackage{bm}
\usepackage{hyperref}
\usepackage{graphicx}
\usepackage{epsfig}
\usepackage{slashed}

\newcommand{\beq}{\begin{equation}}
\newcommand{\eeq}{\end{equation}}
\newcommand{\eq}[1]{Eq.~(\ref{#1})}

\newcommand{\diag}{\mathrm{diag}}

\begin{document}

\title {Pentaquarks with Hidden Charm as Hadroquarkonia}

\author{Michael I.~Eides}
\email[Email address: ]{eides@pa.uky.edu}
\affiliation{Department of Physics and Astronomy, University of Kentucky, Lexington, KY 40506, USA}
\affiliation{Petersburg Nuclear Physics Institute, Gatchina, 188300, St.Petersburg, Russia}
\author{Victor Yu.~Petrov}
\email[Email address: ]{Victor.Petrov@thd.pnpi.spb.ru}
\affiliation{Petersburg Nuclear Physics Institute, Gatchina, 188300, St.Petersburg, Russia}
\author{Maxim V.~Polyakov}
\email[Email address: ]{maxim.polyakov@tp2.ruhr-uni-bochum.de}
\affiliation{Petersburg Nuclear Physics Institute, Gatchina, 188300, St.Petersburg, Russia}
\affiliation{Institut f\"ur Theoretische Physik II, Ruhr-Universit\"at Bochum, D - 44780 Bochum,
Germany}


\begin{abstract}

We consider hidden charm pentaquarks as hadroquarkonium states in a QCD inspired approach.
Pentaquarks arise naturally as bound states of quarkonia excitations and ordinary baryons. The LHCb
$P_c(4450)$ pentaquark  is interpreted as a $\psi'$-nucleon bound state with spin-parity
$J^P=3/2^-$. The partial decay width $\Gamma(P_c(4450)\to J/\psi+N)\approx 11$ MeV is calculated
and turned out to be  in agreement with the experimental data for $P_c(4450)$. The $P_c(4450)$
pentaquark is predicted to be a member of one of the two almost degenerate hidden-charm baryon
octets with spin-parities $J^{P}=1/2^-,3/2^-$. The masses and decay widths of the octet pentaquarks
are calculated. The widths are small and comparable with the width of the $P_c(4450)$ pentaquark,
and the masses of the octet pentaquarks satisfy the Gell-Mann-Okubo relation. Interpretation of
pentaquarks as loosely bound $\Sigma_c\bar D^*$ and  $\Sigma_c^*\bar D^*$ deuteronlike states is
also considered.  We  determine quantum numbers of these bound states and calculate their masses in
the one-pion exchange scenario. The hadroquarkonium and molecular approaches  to exotic hadrons are
compared and the relative advantages and drawbacks of each approach are discussed.

\end{abstract}



\maketitle

\section{Introduction}

Long anticipated heavy hadron states with hidden charm (and/or beauty) finally arrived in the recent
years (see, e.g., the review in \cite{jmr2016} and references therein) and are here to stay.
Four-quark states with hidden charm were discovered first, and the heavy pentaquarks followed
\cite{LHCb2015}.

There are at least four possible scenarios for the dynamics of the LHCb pentaquarks. In the QCD
inspired scenario one assumes that pentaquarks arise as a result of chromoelectric dipole
interaction between a small quarkonium (charmonium) and a large baryon
\cite{dubvol2008,sibvol2005,livol2014,epp2016,PPS} (heavy quarkonium interaction with nuclei was considered in \cite{sjbisgt,lms1992}, see also references in \cite{volsh2008}). Smallness of the quarkonium state is due to large masses of the heavy quarks. The strength of the quarkonium-proton interaction in this case is determined by in principle  calculable quarkonium chromoelectric polarizability and by the proton energy-momentum density. The last one is normalized to the proton mass and is to a large extent model-independent. Pentaquarks in this scenario look like atomlike systems with a small nucleus whose role plays the quarkonium state and light nucleon  quarks that play the role of the atomic electrons. The characteristic feature of this scenario is that the pentaquark decay into a charmonium state with hidden charm and an ordinary baryon is by far the dominant mode of decay. Decays into states with open charm are strongly suppressed because they can go only via exchange by a heavy open charm meson. We will discuss the hadrocharmonium scenario in more detail below.

Molecular-like scenarios initiated in \cite{volok1976} rely on an analogy between heavy exotic
hadrons with hidden charm and molecules. In this scenario charmed constituents of hidden-charm
hadrons preserve their individuality and form bound states. There are two kinds of the
molecular-like scenarios. In the first one charmed hadrons interact via exchange of light mesons and form hidden charmed pentaquarks with the binding energy at the level of hundreds of MeV (see review
in \cite{cclz2016} and references therein). In the second kind of molecular scenario initiated in
\cite{torn1991,teoegk1993,torn1994} the binding energy is at the level of tens of MeV, and the heavy exotic hadrons are bound due to the one-pion exchange. This approach mimics the loosely bound
deuteron. In the deuteron the $S$-wave one-pion central potential is not strong enough to bind the
proton and neutron, and the much stronger noncentral tensor potential does not contribute to the $S$-wave. Binding in the deuteron arises because the tensor potential supports coupling between the $S$- and $D$-states. This mechanism was generalized for the case of the tetraquark mesons with hidden charm in
\cite{torn1991,torn1994}. Below we will develop this approach and apply it to the dynamical
interpretation of the newly discovered pentaquarks. The generic feature of both molecular scenarios
is the necessity to introduce a small distance repulsive cutoff. Technically  this cutoff is needed
to avoid collapse of the would be bound state. Physically, cutoff arises because due to the finite
size both of the constituent hadrons and the exchanged mesons, the potential picture does not work
at small distances. Similar cutoff is routinely introduced in nuclear physics, see, e.g.,
\cite{bertul2007}. The distances between the open charm constituents in the molecular scenario are
relatively large, what  strongly impedes possible decays into final states with hidden-charm mesons
like  $J/\psi$. This seems to be a problem for this scenario, since the LHCb pentaquarks were
discovered as bumps in the invariant mass distributions of $J/\psi N$. We will discuss the role of
the cutoff and other features of the molecular scenario in more detail below.

One more popular idea is to treat heavy exotic hadrons with hidden charm as "true" tetra- and pentaquarks. This approach to heavy pentaquarks with hidden charm was initiated in \cite{mpr2015} and developed further in numerous later publications, see, e.g., \cite{amnss2015,leb2015,lhh2015,hycckc2015,aaar2016} and references in \cite{cclz2016,als2017}. The idea is that the LHCb pentaquarks are diquark-diquark-antiquark bound states. The characteristic feature of this approach is that the hidden-charm pentaquarks arise as compact structures, more or less on par with the ordinary hadrons. The mere assumption about the diquark-diquark-antiquark structure of pentaquarks allows one to develop a rather reach phenomenology. Consideration of the color and flavor assignments for diquarks leads to the prediction of  the flavor pentaquark multiplets \cite{mpr2015}. The multiplet pattern in the diquark-diquark-antiquark is qualitatively different than the one in the hadrocharmonium and molecular approaches and can serve as an experimental signature that allows to choose between different models. The $SU(3)$ flavor symmetry was used to predict ratios of partial weak decay widths of bottom baryons to a pseudoscalar meson and pentaquark \cite{lhh2015,hycckc2015}.  These predictions were further developed in the framework of the effective Hamiltonian approach to pentaquarks in \cite{aaar2016} where the flavor $SU(3)$ symmetry was amended by the heavy quark symmetry. A whole spectrum of new pentaquark states with definite properties arises in this approach. Also a very interesting set selection rules for weak decays of bottom baryons to pentaquarks is predicted. All these results could be used as a guideline in experimental searches for pentaquarks in bottom baryon decays.

In one more scenario pentaquarks  are considered as molecular-like bound states of a "baryon" and a "meson" with an open color \cite{amam2015}. There are also suggestions in the literature that the LHCb results could be explained by some kinematical effects \cite{ugmjao2015,mikh2015,gmwy2015,lreo2016} without need for pentaquarks.

Below we will concentrate on the hadrocharmonium and molecular approaches to pentaquarks. We will describe in detail the interpretation of the LHCb pentaquarks as bound states of charmonium and the nucleon suggested in our previous paper \cite{epp2016}, calculate the pentaquark masses and widths, and predict new pentaquark states. We will also present some new results in molecular  approach and compare our results in hadrocharmonium and molecular approaches with  the predictions of other authors.

\section{Quarkonium-Nucleon Interaction}

Different models were suggested for description of hadrons with hidden charm. Especially appealing
is the hadroquarkonium approach to tetraquarks put forward in \cite{dubvol2008} (see also
\cite{sibvol2005,livol2014}). In \cite{epp2016} we applied this idea to the LHCb pentaquarks.  The
hadroquarkonium approach is based on the simple observation that interaction of a small size heavy
quarkonium with other hadrons can be considered in the framework of the QCD multipole expansion
\cite{kgottf1978,mbv1979,pesk1979,bp1979}, the role of the small parameter plays the ratio of the
quarkonium size and the gluon wave length. For the quarkonium-nucleon interaction this ratio is just
the ratio of the quarkonium and nucleon sizes. In the leading order approximation we have to
consider emission (or absorption) of a chromoelectric dipole gluon by a heavy quark-antiquark pair.
The color singlet pair goes into a color octet state after interaction with a dipole gluon and a
second dipole interaction is needed to return it to the color singlet state. As a result in the
leading approximation interaction of a heavy singlet quark-antiquark pair with other hadrons is
described by the effective Hamiltonian (see, e.g., \cite{volsh2008})

\beq \label{tensintpol} H_{eff}(\bm x)=-\frac{1}{2}\alpha_{ij}E^a_i(\bm x)E_j^a(\bm x), \eeq

\noindent where $E^a_i$ is the chromoelectric field with the absorbed strong coupling
constant\footnote{The gluon part the QCD Lagrangian has the form $-(1/4g_s^2)G^2$.} $\alpha_s$, and
$\alpha_{ij}$ is the quarkonium chromoelectric polarizability

\beq \label{temsorpol} \alpha_{ij}= \frac{1}{16}\langle\psi|(t_1^a-t^a_2)
r_iGr_j(t_1^a-t^a_2)|\psi\rangle, \eeq

\noindent where $t_i^a$ are the $SU(3)_c$ color generators in the fundamental representation, $\bm
r=\bm r_1-\bm r_2$ describes the relative positions of the quark and antiquark, and $G$ is the
quark-antiquark Green function in the color octet channel. In the nonrelativistic approximation the
heavy quark-antiquark  interaction is described by an effective Coulomb potential. Spins and
coordinates decouple  in a nonrelativistic bound state and the effective dipole Hamiltonian for
quarkonium $S$-states reduces to

\beq \label{efffiledham} H_{eff}(\bm x)(S)=-\frac{1}{2}\alpha\bm E^a(\bm x)\cdot\bm E^a(\bm x),
\eeq

\noindent where

\beq \label{polnc} \alpha(nS)=\frac{1}{48}\left\langle nS\left|(t_1^a-t^a_2)\bm rG\bm
r(t_1^a-t^a_2)\right|nS\right\rangle. \eeq

The perturbative quarkonium chromoelectric polarizability for an arbitrary $nl$ states was
calculated long time ago \cite{leut1981,volosh1982} (see also calculations for $nS$ states in the
large $N_c$ limit \cite{pesk1979,bp1979} and the recent calculation of the $1S$ polarizability for
an arbitrary $N_c$ in \cite{bramb2015}). Below we will need polarizabilities for the two lowest
energy levels of nonrelativistic quarkonium \cite{leut1981,volosh1982}

\beq \label{pertpolarizarb} \alpha(1S)=\frac{78}{425}a_0^4m_Q, \qquad
\alpha(2S)=\frac{67264}{663}a_0^4m_Q, \eeq

\noindent where $m_Q$ is the heavy quark mass, and  $a_0=3\alpha_s/4m_q$ is quarkonium Bohr radius.

We will also need the transitional $1S-2S$ polarizability that can be easily calculated in the large
$N_c$ limit along the lines described in \cite{pesk1979,bp1979}

\beq \alpha(1S-2S)=-\frac{3200\sqrt{2}}{6561}a_0^4m_Q. \eeq

\noindent Fitting the $J/\psi$ and $\psi'$ masses, we obtain numerical values for the Coulombic
polarizabilities

\beq
\label{numpolariz}
\alpha(1S)=0.2~\mbox{GeV}^{-3}, \qquad \alpha(2S)=12~\mbox{GeV}^{-3},\qquad
\alpha(1S-2S)=-0.6~\mbox{GeV}^{-3}.
\eeq

\noindent Charmonium is not a Coulombic system and therefore one cannot expect quantitative
agreement between the charmonium polarizabilities and their perturbative values. Transitional
polarizabilities $|\alpha(J/\psi-\psi')|\approx 2$ GeV$^{-3}$ and
$|\alpha(\Upsilon-\Upsilon')|\approx 0.66$ GeV$^{-3}$ for charmonium and bottomonium were extracted
from the phenomenological analysis of the pionic decays $\psi'\to J/\psi\pi\pi$ and
$\Upsilon'\to\Upsilon\pi\pi$ \cite{volosh2004}. Comparing these values with the perturbative results
above we see that the perturbative  calculations provide at best an order-of-magnitude estimates of
the true polarizabilities. Below we will use perturbative polarizabilities for such estimates and
for rough comparison of the relative magnitudes of the polarizabilities.

To describe interaction of a heavy quarkonium with a nucleon we need to calculate the expectation
value of the chromoelectric field squared in \eq{efffiledham} in a nucleon state. To facilitate this calculation we represent the chromoelectric field squared as a linear combination of the
covariant gluon field strength $G_{\mu\nu}^2$ and the energy density of the gluon field (zero
component of the gluon energy-momentum tensor $T^G_{00}$)

\beq
\bm E^2 =\frac{\bm E^2-\bm B^2}{2}+\frac{\bm E^2+\bm B^2}{2}=-\frac{G^2}{4}+g^2T^G_{00},
\eeq

\noindent where $g^2$ is the QCD coupling constant normalized at the scale of the quarkonium radius. It arises here because the QCD coupling constant describing interaction of a small chromoelectric
quark-antiquark dipole with an external field is normalized at the quarkonium size (in our notation
this coupling constant is swallowed by the field).

Exploiting the QCD scale anomaly we obtain

\beq  \label{E2} \bm E^2(\bm x)=g^2\left(\frac{8\pi^2}{bg^2_s}{T^\mu}_\mu(\bm x) + T^G_{00}(\bm
x)\right),
\eeq

\noindent
where  ${T^\mu}_\mu$ is the trace of the QCD energy-momentum tensor, $b=({11}/{3})
N_c-(2/3) N_f$ is the leading coefficient of the $\beta$-function, and $g_s$ is the running strong
coupling constant at the scale of the nucleon radius. Scale dependence of the coupling constant can
be safely ignored for charmonium, but could become important for bottomonium. We temporarily omit
the light quark masses in the trace of the energy-momentum tensor, they will be accounted for
later.

With the help of the representation in \eq{E2} the effective Hamiltonian in \eq{efffiledham} reduces
to the static nonrelativistic potential that describes interaction of a heavy quarkonium with the
nucleon

\beq V(\bm x)=-\frac{1}{2}\alpha g^2\left(\frac{8\pi^2}{bg^2_s}{T^\mu}_\mu(\bm x) + T^G_{00}(\bm
x)\right), \eeq

\noindent where $T^{\mu}_\mu(\bm x)$ and $T^G_{00}(\bm x)$ are the respective tensor densities
inside the proton. The energy density $T^G_{00}(\bm x)$ carried by the gluons inside the proton
cannot be determined unambiguously and is model-dependent. We make a natural assumption that it is
proportional to the total proton energy density $T^G_{00}(\bm x)=\xi T_{00}(\bm x)$ \cite{ns1981}.
Such assumption worked pretty well in the case of the pion. The factor $\xi$ depends on the
normalization point and is about $1/2$ at $Q^2\sim 1$ GeV$^2$ \cite{ns1981}. We will assume this
value in further calculations. It is convenient to represent the interaction potential in terms of
the energy density and pressure, $\diag{\left(T_{\mu\nu}(\bm x)\right)}=(\rho_E(\bm x),p(\bm x),p(\bm
x),p(\bm x))$

\beq \label{effpot} V(\bm{x}) = -\alpha\frac{4\pi^2}{b }\left(\frac{g^2}{g_s^2}\right)\left[
\rho_E(\bm x) \left(1+\xi\frac{bg^2_s}{8\pi^2}\right) - 3 p(\bm x)\right]. \eeq

\noindent This effective potential has a simple interpretation. A point-like quarkonium serves as a
tool that scans the local energy density and local pressure inside the nucleon. It could happen that
the size of quarkonium is not small enough in comparison with the size of the nucleon. In such case
we will need to consider higher order terms in the QCD multipole expansion in order to improve
description of the quarkonium-nucleon interaction.

The potential in \eq{effpot} is to a large extent model-independent, its normalization

\beq
\label{intV}
\int d^3x V(\bm{x}) = -\alpha \frac{4\pi^2}{b }\left(\frac{g^2}{g_s^2}\right){
M}_N \left(1+\xi\frac{bg_s^2}{8\pi^2}\right)
\eeq

\noindent
is determined by the total energy of the nucleon $\int d^3 x \rho_E (\bm x)=M_N$ and the
stability condition $\int d^3x p(\bm x)=0$. Only the factor $\nu= 1+\xi({bg_s^2}/{8\pi^2})$ in
\eq{intV} cannot be determined from the first principles, and we use the phenomenological value
$\xi\sim1/2$ to obtain $\nu\sim 1.5$.

We are going to use the interaction potential in \eq{effpot} to explore possible bound states formed by heavy quarkonia states and the nucleon. With the known normalization of the potential and the nucleon radius we could proceed in an almost model-independent way\footnote{A model-independent estimate of the minimal polarizability sufficient for existence of a bound state was also obtained in \cite{PPS}.}, considering the potential as a potential well with the size of the nucleon. Instead we will use the  local energy density $\rho_E(\bm x)$ and pressure $p(\bm x)$ that were computed in the $\chi$QSM in  \cite{maxim2007}. The potential constructed in this way automatically satisfies the normalization  condition in \eq{intV} since the normalization condition for the energy density and the stability  condition for the pressure hold due to equations of motion.

\section{Pentaquarks as Hadrocharmonium States}

The LHCb pentaquarks were discovered in the analysis of the invariant mass distributions of $J/\psi$
plus nucleon. A natural idea is that the pentaquarks arise as bound states of charmonium excitations
and the nucleon. We use the nonrelativistic quarkonium-nucleon potential in \eq{effpot} to explore
this hypothesis. The nonrelativistic Schr\"odinger equation in the channels  $J/\psi+ N$ and $\psi'+
N$   has the form

\beq \label{Sch1} \left(-\frac{\bm\nabla^2}{2\mu}+ V(r)-E_b\right)\psi_b  = 0, \eeq

\noindent where $\mu$, $\psi_b$ and $E_b$ are the reduced mass, wave function, and the binding
energy, respectively. The chromoelectric polarizabilities for each channel are collected in
\eq{numpolariz}. Due to the poor knowledge of polarizabilities we will vary them in a relatively
wide region.

Solving the Schr\"odinger equation \eq{Sch1} numerically we find that a bound state of $J/\psi$ and
the nucleon arises only when the polarizability reaches the critical value $\alpha=5.6$ GeV$^{-3}$.
This value of polarizability is more than an order of magnitude larger then the perturbative
$\alpha(1S)$ in \eq{numpolariz}, and we conclude that $J/\psi$ does not form a bound state with the
nucleon. Critical values of polarizabilities for the excited $S$ states of charmonia are far below
their perturbative values in \eq{numpolariz} (see \cite{leut1981,volosh1982,pesk1979,bp1979,epp2016}
for the perturbative polarizabilities of the higher excited states of charmonia). Hence, such states do form bound states with the nucleon\footnote{To the best of our knowledge stronger binding between $\psi'$ and nuclei was first mentioned in \cite{lms1992}.}. We concentrate on the lowest $\psi'N$ bound states in this paper.

A bound $\psi'N$ state with the mass of the $P_c(4450)$ pentaquark, the binding energy
$E_b=-176$~MeV, and the orbital momentum $l=0$ is formed at $\alpha(2S)=17.2$~GeV$^{-3}$. This value
of polarizability is well inside the error bars of the perturbative calculation of the $\alpha(2S)$
polarizability in \eq{numpolariz}. There are no other bound $\psi'N$ states at
$\alpha(2S)=17.2$~GeV$^{-3}$.

Exploiting the uncertainty in our knowledge of the $\alpha(2S)$ polarizability we can also adjust it
in such way as to match the light LHCb pentaquark. A bound $\psi'N$ state with the mass of the
$P_c(4380)$ pentaquark, the binding energy $E_b=-246$~MeV, and the orbital momentum $l=0$ is formed
at $\alpha(2S)=20.2$~GeV$^{-3}$. Again, there are no other bound $\psi'N$ states at this value of
polarizability. An identification of this bound state with the $P_c(4380)$ pentaquark would mean
that there  are no heavier pentaquarks formed by $\psi'N$.

Taking into account opposite parities of the observed LHCb pentaquarks \cite{LHCb2015} it is
interesting to explore possible $\psi'N$ bound states with $l=1$. Such state arises for the first time
when polarizability reaches the value $\alpha\approx 22.4$~ GeV$^{-3}$. One could try to identify
this state together with a more tightly bound $l=0$ bound state with the pair of the LHCb
pentaquarks. The spin-parities ${3}/{2}^-$ for the lighter state and ${5}/{2}^+$ for the heavier one
fit nicely the experimental data. However, the mass splitting between these states is about 300 MeV
instead of the experimentally observed 70 MeV. The large mass difference between the rotational
excitation and the ground state indicates that the moment of inertia of the bound state is small.
This bound state moment of inertia is determined by the size of the binding potential. The
binding potential is proportional to the nucleon energy density, and hence the same binding
potential determines the nucleon moment of inertia.  In the mean field picture of the nucleon its
moment of inertia determines the energy of its rotational excitations which is about a few hundred
MeV as can be seen from the $N-\Delta$ mass splitting. Due to the connection
between the nucleon moment of inertia and the bound state moment of inertia we are compelled
to conclude that the moment of inertia of the bound state is small. This explains large splitting
between the bound states with different angular momenta. Another drawback of the scenario with two pentaquarks as $l=0$ and $l=1$ bound states is that it predicts that the heavier  pentaquark
with $l=1$ has a larger decay width, what squarely contradicts the experimental data in
\cite{LHCb2015}. Both due to the prediction of a too large mass splitting and an unrealistic hierarchy
of decay widths we reject the interpretation of the LHCb pentaquarks as $l=0$ and $l=1$ bound states
of $\psi'N$.

The hadrocharmonium interpretation of pentaquarks was tested in \cite{PPS} in the framework of the  Skyrme model. It turned out that the Skyrme model energy-momentum tensor densities lead to the same  conclusions as the considerations in  \cite{epp2016} which were loosely based on the chiral quark  soliton model. This demonstrates that the hadrocharmonium interpretation of  pentaquarks is robust  and does not depend on the details of a particular nucleon model. New hadrocharmonium  $\psi'\Delta$ bound states with hidden charm, isospin $3/2$ and masses $4.5$~GeV and $4.9$~GeV were predicted in \cite{PPS}.

In summary, solving the Schr\"odinger equation we have found two theoretically acceptable values of
polarizability that admit interpretation of either of the two LHCb pentaquarks as a $\psi'N$ bound
state. Only one bound state exists at each value of polarizability, and, respectively, only one of
the observed pentaquarks can be interpreted as a $\psi'N$     bound state. Experimentally the
$P_c(4380)$ peak has a rather large width $205\pm18\pm 86$~MeV, whereas the $P_c(4450)$ peak is
narrow with the width $39\pm5\pm 19$~MeV. To make a choice between the two possible hadrocharmonium
interpretations of the LHCb pentaquarks we need to calculate the theoretical decay widths of the
bound state solutions found above.

\section{Partial width of the $\psi'N$ bound state}

Interaction of heavy charmonia states with the nucleon is described by the nonrelativistic potential
in \eq{effpot}. This potential is universal, only its overall strength, determined by the
polarizability of the respective charmonia excitation,  changes when we go from one charmonia state
to another. The nonzero transition polarizability in \eq{numpolariz} shows that there exists a
similar nondiagonal potential that describes the transition $J/\psi\to\psi'$ off the nucleon. Due to
the coupling between the $J/\psi N$ and $\psi'N$ channels the pentaquark that we found in the
$\psi'N$ channel should arise as a resonance in the $J/\psi N$ scattering channel. We are going to
solve the two-channel scattering problem, find the resonance $J/\psi N\to J/\psi N $ scattering
amplitude, and determine the width of the resonance by comparing this scattering amplitude with the
standard Breit-Wigner expression.

The Hamiltonian for the two-channel nonrelativistic scattering problem has the form

\beq \label{2chham} H=\left(
\begin{array}{c|c}
-\frac{\nabla^2}{2\mu_1}+V_{11}&V_{12}\\ \hline V_{12}&-\frac{\nabla^2}{2\mu_2}+V_{22}+\Delta
\end{array}
\right), \eeq

\noindent where $\Delta=m_{\psi'}-m_{J\psi}$, $\mu_1=m_{J/\psi} m_N/(m_{J/\psi}+ m_N)$,
$\mu_2=m_{\psi'} m_N/(m_\psi+ m_N)$. The potentials $V_{ij}$ are obtained from the potential in
\eq{effpot} by substituting the respective polarizabilities from \eq{numpolariz} instead of
$\alpha$.

Next we solve the scattering problem for the Schr\"odnger equation

\beq \label{2channel} H\Psi=E\Psi, \qquad \Psi=\left(
\begin{array}{c}
\psi_1\\ \psi_2
\end{array}
\right), \eeq

\noindent where $E$ is the nonrelativistic $J/\psi N$ energy in the center of mass frame ($E=\bm
q^2/2\mu_1$, $\bm q$ is the relative momentum) with only the incoming plane wave $\psi_1=e^{i\bm
q\cdot\bm x}$ in the $J/\psi N$ channel different from zero.

The transition potential $V_{12}$ is small and the perturbation theory treatment is sufficient for the scattering problem in \eq{2channel}. Due to coupling between the channels the incoming plane wave $\psi_1(\bm x)$ leaks in the $\psi'N$ channel

\beq \label{leakkf} \psi_{2}({\bm x})=-\int d^3x^{\prime}G_2({\bm x},{\bm x'})V_{12}({\bm
x}^{\prime})e^{i {\bm q}\cdot{\bm x}^{\prime}}.
\eeq

\noindent Here

\beq
G_2({\bm x},{\bm x'})=\left\langle {\bm x}\left|\frac{1}{-\frac
{\bm\nabla^{2}}{2\mu_2}-E+\Delta+V_{22}-i0}\right|\bm{x}^{\prime}\right\rangle
\eeq

\noindent is the Green function in the $\psi'N$ channel (see \eq{2chham} and  \eq{2channel}). Near
the resonance

\[
G_2({\bm x},{\bm x'}) = \frac{\psi_{R}({\bm x})\psi_{R}^*({\bm x'})}{E_R-E},
\]

\noindent where $E_R$ is the resonance energy. The wave function  $\psi_{2}({\bm x})$ in \eq{leakkf}
in its turn generates correction to the incoming plane wave $\psi_{1}({\bm x})$, that near the
resonance has the form

\beq
\label{corr2}
\delta\psi_1(\bm x) = \int d^3 x' G_1({\bm x},{\bm x'})V_{12}({\bm
x}')\psi_R({\bm x}') \frac{\int d^3 x'' V_{12}({\bm x}'')\psi^*_R({\bm x}'')e^{i {\bm q}\cdot{\bm
x}^{\prime\prime}}}{E_R-E},
\eeq

\noindent where

\beq
G_1(\bm x ,\bm x')=\left\langle\bm x\left|\frac{1}{-\frac{\bm\nabla^2}{2\mu_1}-E-i0}\right|\bm
x'\right\rangle =2\mu_1\frac{e^{iq|\bm x-\bm x'|}}{4\pi|\bm x-\bm x'|}
\eeq

\noindent is the free Green function in the $J/\psi N$ channel.

Calculating $\delta\psi_1(\bm x)$ near the resonance with the orbital momentum $l$ at
large  $r=|\bm x|$ we obtain the wave function in the $J/\psi N$ channel as a superposition of the
incoming plane wave and the outgoing spherical wave

\beq
\label{ouraympt}
\psi_1(\bm r)+\delta\psi_1(\bm r) =e^{i\bm q\cdot\bm
x}+2\mu_1\frac{e^{iqr}}{r}\frac{1}{E_R-E}(2l+1)P_l(\cos\theta)\left|\int_0^\infty dr
r^2j_l(qr)R_{l}(r)V_{12}(r)\right|^2,
\eeq

\noindent
where the resonance radial wave function $R_l(r)$ is normalized by the condition
$\int_0^\infty dr r^2R_l^2(r)=1$.

This wave function can be written in terms of the scattering amplitude  $f(\theta)$ ($\theta$ is the
scattering angle)

\beq
\label{scatamp}
\psi_1({\bm x})+\delta\psi_1(x) = e^{i{\bm q}\cdot {\bm x}} + f(\theta)
\frac{e^{iqr}}{r}.
\eeq

\noindent The scattering amplitude near the resonance has the standard Breit-Wigner form

\beq f(\theta)=-\frac{2l+1}{q}\frac{\Gamma/2}{E-E_R} P_l(\cos\theta), \eeq

\noindent where $\Gamma$ is the resonance partial decay width in the $J/\psi N$ channel.

Comparing \eq{ouraympt} and  \eq{scatamp} we obtain

\beq \Gamma \label{partwipsi} =4\mu_1 q\left|\int_0^\infty dr r^2 R_l(r)V(r) j_l(q r)\right|^2,
\eeq

\noindent where $q=\sqrt{2\mu_1E_R}$ and $j_l(z)$ is the spherical Bessel function.

\section{Phenomenology of charmonium-baryon bound states}

We obtained above a candidate for the heavy LHCb pentaquark as a bound $\psi'N$  state that arises
at $\alpha(2S)=17.2$~GeV$^{-3}$, and a candidate for the light LHCb pentaquark as a bound $\psi'N$
state that arises at $\alpha(2S)=20.2$~GeV$^{-3}$. Now we are in a position to calculate these bound
state partial decay widths into the $\psi N$ channel. Using the phenomenological value of the
transitional polarizability $\alpha(2S\to 1S)=2$~GeV$^{-3}$ \cite{volosh2004} we obtain  partial widths at the level of tens of MeV for both
bound states.  We also made a rough estimate of the
partial width for the decay of either of the $\psi'N$ bound states into $J/\psi+N+\pi$, and it
turned out to be even smaller than the partial width into the $J/\psi+ N$ channel. The decays of
these bound states into (anti)charmed meson + charmed baryon are strongly suppressed in this
scenario, since such decays into open charm channels can go only via $t$-channel exchange by a heavy
$D$-meson. Therefore the total width of each of the $\psi'N$ bound states is small, in the range of
tens of MeV.

The LHCb $P_c(4380)$ pentaquark is a wide peak with the width $205\pm18\pm 86$~MeV, while the
$P_c(4450)$ pentaquark is a narrow state with the total width $39\pm5\pm 19$~MeV. The acceptable
spin-parity assignments include $(3/2^-,5/2^+)$, $(3/2^+,5/2^-)$, and $(5/2^+,3/2^-)$, all with
opposite parities \cite{LHCb2015}. Comparing the experimental data with the calculations above we
interpret the narrow heavy $P_c(4450)$ pentaquark as the $\psi'N$ bound state. This $\psi'N(4450)$
bound state is formed in the $S$-wave and is a $J^P=3/2^-$ state. We used \eq{partwipsi} to
calculate its partial width and,  in reasonable agreement with the data, obtained
$\Gamma(P_c(4450)\to N+J/\psi)\approx11$~MeV for the dominant decay mode.

The potentials in \eq{Sch1} are spin-independent, so there are two degenerate bound states with
$J^P=1/2^-$ and $J^P=3/2^-$. The hyperfine splitting between these degenerate  color-singlet bound
states arises due to interference of the chromoelectric dipole $E1$ and the chromomagnetic
quadrupole $M2$ transitions in charmonium. It can be described by the effective Hamiltonian

\beq H_{eff} =-\frac{\alpha}{4m_Q}S_j\langle N|[E^a_i(D_iB_j)^a+(D_iB_j)^aE^a_i]N\rangle, \eeq

\noindent where $S_j$ is the quarkonium spin, $\alpha$ and $m_Q$ are the same chromoelectric
polarizability and the heavy quark mass as above, and only the nucleon matrix element of the product
of chromoelectric and chromomagnetic fields requires calculation.

The strength of this interaction is determined by the chromoelectric polarizability and it is
additionally suppressed by the heavy quark mass $\sim 1/m_Q$.  A semiquantitative estimate of the
hyperfine splitting produces a small value in the range of $5-10$~MeV. Therefore we expect to find
two almost degenerate pentaquark states with $J^P=1/2^-$ and $J^P=3/2^-$ and with the mass of the
observed pentaquark $4450$~MeV. It would be very interesting if the LHCb collaboration could check
this hypothesis in their partial wave analysis.

Thus far we ignored flavor symmetry of ordinary baryons. Recall that  the nucleon is a member of the
baryon octet. The interaction potential in \eq{effpot} is proportional to the matrix element of $\bm
E^2$, and in the linear  approximation in the quark mass it is one and the same for all members of the
baryon octet. Therefore we should expect that all members of the baryon octet bind with $\psi'$, and
the respective  pentaquarks also form an octet. Masses of these octet pentaquarks are just the sums
of the constituent masses and the binding energies. The binding energy depends on the mass of the
ordinary octet baryon $B$ only through the reduced mass in the kinetic energy in the respective
\eq{Sch1}. Then the pentaquark octet mass splittings in the leading order in the ordinary octet mass splitting $\Delta M$ are (see the definition of the reduced mass $\mu_1$
after \eq{2chham})

\beq  \label{massploctpen}
\Delta E=-\frac{\mu_1}{m_N^2}\left\langle
N\left|-\frac{\bm\nabla^2}{2\mu_1}\right|N\right\rangle\Delta M.
\eeq

\noindent
We checked this result by solving \eq{Sch1} for each ordinary octet baryon and calculating the pentaquark binding energies (and their changes) directly. Both approaches lead to the same results.

We have also solved the two-channel scattering problems for $J/\psi$ scattering off all members $B$ of the ordinary baryon octet, found the respective $\psi'B$ resonances,  and calculated their partial decay widths  into $J/\psi B$. The results for the octet pentaquarks mass splittings and widths are collected in Table~\ref{pentoct}. All octet pentaquarks $P_B$ have very small decay widths into $J/\psi+B$. We expect that like $P_c(4450)$ they also have small total widths. The mass splittings between the octet pentaquarks in Table~\ref{pentoct} are somewhat smaller then the mass splitting in the ordinary baryon octet as predicted by \eq{massploctpen}.  Due to hyperfine splitting there are two almost degenerate pentaquark octets with $J^P=1/2^-$ and $J^P=3/2^-$. With very good accuracy the states in the pentaquark octets satisfy the Gell-Mann-Okubo mass formula

\beq \frac{m_{P_N}+m_{P_\Xi}}{2}=\frac{m_{P_\Sigma}+3m_{P_\Lambda}}{4}, \eeq

\noindent or, numerically, $4613$ MeV$\approx4615$ MeV.

\begin{table}[h!]
\caption{\label{pentoct} Penta Octet ($J^P=3/2^-$): Masses and Widths}
\begin{ruledtabular}
\begin{tabular}
{lcccc} $P_B$\footnote{Penta octet states, $P_B$ is a $\psi'B$ bound state.} & Mass (MeV) &
$M_P-M_{P_c}$ (MeV)\footnote{Mass differences between the penta octet states and $P_c$.} & $M_B-M_N$
(MeV)\footnote{Mass differences between the baryon octet states and $m_N$.}&  Width
(MeV)\footnote{Partial width for decays of penta octet states into $J/\psi+B$}
\\
\hline
$P_N$ ($P_c(4450)$)& 4449 & 0 &0& 11
\\
$P_\Sigma$&4665&217& 253&14
\\
$P_\Lambda$&4598&150& 176&13
\\
$P_\Xi$&4776&327& 378 &15
\\
\end{tabular}
\end{ruledtabular}
\end{table}

Only the heavy narrow pentaquark with spin-parity $3/2^-$ finds a natural interpretation as a $\psi'N$ bound state in the scenario described above. Another explanation should be found for the wide $P_c(4380)$
pentaquark\footnote{Let us mention the suggestions in the literature that there is really no
resonance at the position of this pentaquark, see. e.g., \cite{lreo2016}.} with spin-parity $5/2^+$. The hadrocharmonium approach predicts also bound states formed by the nucleon and other excited states of charmonium, besides $\psi'$. Scanning the charmonium  spectrum in search of a state that could bind with the nucleon to form a pentaquark with mass $4380$ MeV and spin-parity $5/2^+$ we observe that $\chi_{c2}(3556)$ with $J^P=2^+$ has necessary quantum numbers and mass.

To find out if  $\chi_{c2}(3556)$ really binds with the nucleon we have to solve a dynamical
problem. As a nonrelativistic heavy quark-antiquark excitation $\chi_{c2}(3556)$ is a $P$-wave
state. The chromoelectric polarizability of this $P$-state is a two-index symmetric tensor
$\alpha_{ij}$ that can be calculated using \eq{temsorpol}. Then one can calculate the
$\chi_{c2}(3556)$-nucleon interaction  potential starting with \eq{tensintpol}, like it was done above
in the case of the $S$-wave charmonium states, and solve the respective bound state Schr\"odinger
equation. An estimate of the perturbative polarizability tensor for the $P$-state shows that it has
roughly the same magnitude as for the $S$-state. Thus we have every reason to expect that
$\chi_{c2}(3556)$ forms a bound state with the nucleon with spin-parity $5/2^+$ and mass about
$4380$ MeV. This state could be a candidate for the observed $P_c(4380)$ pentaquark. Moreover, due to the
smallness of the spin-spin interaction there should be also almost degenerate states with
spin-parities $1/2^+,3/2^+$. These particles do not exhaust the reach spectrum of pentaquarks formed
by the $P$-states of charmonia  and the nucleon. We expect to find $\chi_{c0}$-nucleon bound state
with spin-parity  $1/2^+$, almost degenerate $\chi_{c1}$-nucleon bound states with spin-parities
$1/2^+,3/2^+$, and almost degenerate $h_c$-nucleon bound states with spin-parities  $1/2^+,3/2^+$.
All these states should be very narrow because there are no open channels for decays except decays
into particles without hidden charm that are strongly suppressed in accordance with the
Okubo-Zweig-Iizuka rule. Suppression of strong decays of these particles puts a question mark over
the possibility to identify the $\chi_{c2}(3556)$-nucleon bound state with the LHCb $P_c(4380)$
pentaquark.

\section{Are there Bottomonium-nucleon hadrocharmonium bound states?}

The binding mechanism developed above could generate bound states of bottomonium and the nucleon.
The perturbative polarizability in \eq{pertpolarizarb} (see also
\cite{leut1981,volosh1982,pesk1979,bp1979}) depends on  the heavy quark mass, the running strong
coupling constant, and the Bohr radius. Both the effective coupling constant and the Bohr radius for
bottomonium are smaller  than for charmonium, while the mass of the bottom quark is larger than the
mass of the charmed quark. Taking into account the interplay of these effects we calculated
perturbative polarizabilities for the lowest  states of bottomonia

\beq \label{pertpolarizarbott} \alpha(1S)\approx0.07~\mbox{GeV}^{-3}, \qquad \alpha(2S)\approx
5~\mbox{GeV}^{-3}, \eeq

\noindent where we used the bottom quark mass $m_b=5105$ MeV, the bottomonium Bohr radius
$a_0=3\alpha_s/4m_b\approx0.1$ fm, and the strong coupling  $\alpha_s(a_0)\approx 0.6$ in this
calculations.

We searched for bottomonia-nucleon bound states, solving the Schr\"odinger equation \eq{Sch1} for
bottomonium. No $\Upsilon(1S)$-nucleon bound state was found for a reasonable value of
polarizability. The results for the $\Upsilon(2S)$-nucleon system are inconclusive due to poor
knowledge of polarizability. A bound state could exist but a much better handle on polarizability is
needed to make a definite statement. The sizes of higher bottomonia excitations are comparable to
the size of the nucleon and the dipole approximation at the root of our approach is not valid any
more.

\section{Meson Exchanges, Nuclei, and the Deuteron}

Molecular bound states of two charmed mesons as an explanation of certain states in the charmonium
spectrum were suggested long time ago \cite{volok1976}. Nowadays molecular models of mesons and
baryons with hidden charm are very popular, and there are numerous papers discussing this scenario,
see, e.g., review in  \cite{cclz2016}. We would like to compare the characteristic properties of
pentaquarks arising in the molecular approach with the properties of the hadrocharmonium
pentaquarks. Binding potential in the molecular  approach is due to the exchange of light mesons. It
is modeled after the similar approach widely accepted in nuclear physics. Let us recall the basics
of the meson picture of nuclear forces \cite{bertul2007}.

The forces between nucleons in nuclei are due to exchanges of light mesons: pions, $\eta$-,
$\sigma$-, $\rho$- and $\omega$-mesons. Together they generate the potential resembling the Van der
Waals molecular potential. Attraction at large distances ($\geq1$ fm) is due to the light pion,
attraction at intermediate distances is described by $\sigma$-meson (or two-pion) exchange and the
short distance repulsion is usually ascribed to the $\rho$- and $\omega$-meson exchanges. The meson
exchanges  do not make sense when distances between the nucleons in nuclei become comparable to the
sizes of the exchanged bosons and/or sizes of the constituents. A strong repulsion core at small
distances is usually introduced in the potential. Its position is determined by the  particle sizes
and should be chosen in the range of 0.3-0.5 fm. This approach provides at least a qualitative
description of nuclei. The nucleons in a typical nucleus are separated by the distances about 0.7
fm. At these distances the contribution  of the light pion exchange to the interaction potential is
strongly suppressed and is almost irrelevant in comparison with the $\sigma,\rho,\omega$
contributions.

This is not the case for the loosely bound deuteron. The binding energy in the deuteron is very
small, about 2.2 MeV, and the nucleons in the deuteron are separated by a relative distance about 2
fm. At such distances  only the light pion contribution to the potential survives. Calculating the
nucleon-nucleon scattering amplitude in the nonrelativistic approximation we obtain the momentum
space nucleon-nucleon potential

\beq \label{momsppot} V(\bm q)=-\frac{4g_{\pi NN}^2}{M_N^2}(\bm T_1\cdot\bm T_2)\frac{(\bm S_1\cdot
\bm q)(\bm S_2\cdot \bm q)}{\bm q^2+m_\pi^2}, \eeq

\noindent where $g_{\pi NN}=13.7$ is the pseudoscalar nucleon-pion coupling constant, $m_\pi$ and
$M_N$ are the pion and nucleon masses, and $\bm S_i$ and $\bm T_a$ are the nucleon spin and isospin
operators, respectively.

The potential in  coordinate space  is a sum of a central spin-spin potential and a tensor
potential

\beq \label{onepionnucl} V(\bm r)=V_C(\bm r)+S_{12}(\bm S_1,\bm S_2,\bm n)V_T(\bm r), \eeq

\noindent where $S_{12}(\bm S_1,\bm S_2,\bm n)=3(\bm S_1\cdot\bm n)(\bm S_2\cdot\bm n)-(\bm
S_1\cdot\bm S_2)$, and  formally

\beq \label{centradeutp}
\begin{split}
V_C(\bm r) =&\frac{g_{\pi NN}^2}{M_N^2}(\bm T_1\cdot\bm T_2)(\bm S_1\cdot\bm
S_2)\biggl(m_\pi^2\frac{e^{-m_\pi r}}{3\pi r}-\frac{4}{3}\delta^{(3)}(\bm r)\biggr),\\ V_T(\bm
r)=&\frac{g_{\pi NN}^2}{M_N^2}(\bm T_1\cdot\bm T_2)\left(m_\pi^2r^2+3m_\pi r+3\right)\frac{e^{-m_\pi
r}}{3\pi r^3}.
\end{split}
\eeq

Naively, one could hope that this one-pion exchange potential would be sufficient to describe the
deuteron. This does not happen due to the problems at small distances. Both the $\delta$-function
contribution to the spin-spin potential and the singular $1/r^3$ contribution to the tensor
potential are unphysical, they arise from distances where the one-pion exchange makes no sense due
to finite sizes of all particles. To get rid of unphysical short distance contributions one could
introduce soft or hard core at small distances. Instead it is routine in nuclear physics to
regularize the potential at small distances by inserting the dipole form factor
$[(\Lambda^2-m_\pi^2)/(\Lambda^2+\bm q^2)]^2$ in \eq{momsppot}, see, e.g., \cite{bertul2007}. Naive
insertion of this form factor smears the $\delta$-function contribution in  \eq{centradeutp}. Notice
that the $\delta$-function contribution to the spin-spin potential has the sign opposite to the sign
of the Yukawa type contribution in \eq{centradeutp}. After such formal regularization a repulsive at
large distances spin-spin potential turns into strong attraction at distances shorter than than the
inverse regularization parameter. We consider this regularized $\delta$-function contribution to the
potential unphysical, and subtract it from the regularized potential. Then the regularized
potentials in \eq{centradeutp} have the form (compare \cite{torn1994,tc2008})

\beq \label{regdetpot}
\begin{split}
V_{C,reg}(\bm r) =&\frac{g_{\pi NN}^2}{M_N^2}(\bm T_1\cdot\bm T_2)(\bm S_1\cdot\bm
S_2)\frac{m_\pi^2}{3\pi}Y(\Lambda,m_\pi,r),\\ V_{T,reg}(\bm r)=&\frac{g_{\pi NN}^2}{M_N^2}(\bm
T_1\cdot\bm T_2)\frac{1}{3\pi}Z(\Lambda,m_\pi,r),
\end{split}
\eeq

\noindent where

\beq \label{defymlr} Y(\Lambda,m_\pi,r)=\frac{e^{-m_\pi r}-e^{-\Lambda
r}}{r}-\frac{\Lambda^2-m_\pi^2}{2\Lambda }e^{-\Lambda r}, \quad Z(\Lambda,m_\pi,r)=r\frac{\partial
}{\partial r}\left(\frac{1}{r}\frac{\partial }{\partial r}Y(\Lambda,m_\pi,r)\right). \eeq

\noindent The functions $Y(\Lambda,m_\pi,r)$ and $Z(\Lambda,m_\pi,r)$ are nonsingular and positive
(or negative) definite at all distances. As a result the regularized potentials are finite at zero
and a repulsive (attractive)  at large distances potential remains repulsive (attractive) at all
distances.

It turns out that the one-pion exchange allows quantitative description of the principal deuteron
characteristics with any short distance modifications we just described \cite{torn1991,torn1994}.
There is a nontrivial mechanism at work. The attractive spin-spin potential vanishes in the chiral
limit and is therefore suppressed by the factor $m_\pi^2/M_N^2$. It is not strong enough to bind the
proton and neutron if the tensor potential is turned off. The tensor potential is nonzero even in
the chiral limit and is thus much stronger. It couples $S$- and $D$-waves in the Schr\"odinger
equation and due to this coupling a loosely bound deuteron arises. We obtain the experimental
binding energy 2.2 MeV if we place an infinite wall at $r_0=0.485$ fm. The fraction of the $D$-wave
squared in this case is about 7\% and the deuteron root mean square (rms) radius is 1.98 fm. The
regularized potentials in \eq{regdetpot} reproduce the same binding energy at $\Lambda=800$ MeV, see
also \cite{torn1994,tc2008,lzah2016}. With the regularized potentials the fraction of the $D$-wave
squared is about 5\% and the deuteron rms is 1.92 fm. In both cases the cutoff parameters have a
reasonable magnitude, confining the one-pion potential to distances larger than 0.25-0.4 fm.

The one-pion exchange mechanism modified at small distances completely describes all possible
nucleon-nucleon bound states. There are four different spin-isospin states of two nucleons with
$S=0,1$, $T=0,1$. There is no bound state with $S=T=0$ since the spin-spin potential in this case is
repulsive and the tensor potential turns into zero. The spin-spin potential is attractive and the
tensor one is nonzero in the state with the quantum numbers of the deuteron, $S=1$, $T=0$. The
spin-spin potential for $S=0$, $T=1$ is attractive and coincides with the deuteron spin-spin
potential, while the tensor potential in this case is zero. Finally, in the state with $S=T=1$ the
signs of both potentials are opposite to the deuteron potentials and are suppressed by the factor 3.
The Schr\"odinger equation with the potential in \eq{onepionnucl}  and any short range
regularization discussed above has a bound state solution with the deuteron quantum numbers and the
binding energy 2.2 MeV. There are no bound states with any other spin-isospin quantum  numbers.
There are also no loosely bound nucleon-antinucleon states. One-pion exchange potential changes the
overall sign when one replaces one of the nucleons by an antinucleon. Then absence of the
nucleon-antinucleon states follows from the previous analysis.

Modern nuclear-type potentials (see, e.g., \cite{bertul2007}) are much more sophisticated than the
primitive one-pion exchange and include exchanges by other light bosons. We do not need them for our
discussion of the deuteron. As we have seen the main features of a loosely bound deuteron are due to
the long distance part of the one-pion exchange potential. Quantitative description of the deuteron
requires some kind of short distance cutoff whether we include exchanges by other mesons besides
pion or not, and the nature of this cutoff is only obscured by other mesons.

Below we will consider applications of the one-pion exchange and light boson exchange mechanisms to
the LHCb pentaquarks.

\section{One-Pion Exchange and Pentaquarks}

The one-pion exchange mechanism was generalized for description of tetraquarks with hidden charm in
\cite{torn1991,teoegk1993,torn1994}. The essence of this approach to the loosely bound tetraquarks
is the interplay between the channels with different orbital momenta. With reasonable assumptions
about the magnitude of the short distance cutoff new tetraquark states were predicted in this
framework  \cite{torn1994}. The state $X(3782)$ that was discovered many years later \cite{skcb2003}
turned out to be one of these predicted states \cite{torn2004}. It is worthwhile to figure out if
this nice mechanism could be applied for the description of pentaquarks, can we construct the LHCb
pentaquarks from a charmed baryon and an anticharmed boson  as a deuteronlike loosely bound states.
Inspection of the charmed hadron spectrum shows that the sum of masses of  $\Sigma_c^*(2520)$ and
$\bar D(1870)$ exceeds  the mass of $P_c(4380)$  only by 10 MeV, and the sum of masses of
$\Sigma_c(2455)$ and $\bar D^*(2010)$  exceeds the mass of $P_c(4450)$ only by 15 MeV. This
immediately suggests that the respective LHCb  pentaquarks are  deuteronlike loosely bound states.
At first glance the one-pion exchange mechanism has a fair chance to support the necessary binding
in both cases. However, one-pion exchange cannot bind $\Sigma_c^*\bar D$ because the $\pi\pi D$
vertex is banned by parity conservation. Therefore $P_c(4380)$ cannot be considered as a loosely
bound deuteronlike $\Sigma_c^*\bar D$ state\footnote{See, however, \cite{ssh2016}, where the
$P_c(4380)$ state was interpreted with the help of one-pion exchange for the coupled channels
$\Sigma^*_c\bar D-\Sigma_c\bar D^*$.}. It remains to figure out if $P_c(4450)$ could be interpreted
as a deuteronlike loosely bound  $\Sigma_c\bar D^*$  state.

Let us first obtain the momentum space one-pion potential for interaction of two
arbitrary hadrons. The Goldberger-Treiman relationship $g^N_A/F_\pi=g_{\pi NN}/M_N$ allows to
replace the coupling constant $g_{\pi NN}$ in \eq{momsppot}  by the nucleon axial charge $g^N_A$ and
the pion decay constant $F_\pi=92$ MeV. The coordinate space nucleon-nucleon potential in terms
of the axial charge is obtained from the expressions in \eq{centradeutp}, \eq{regdetpot} by the
substitution $g_{\pi NN}^2/M_N^2\to (g_A^N)^2/F_\pi^2$. The respective hadron-hadron potential is
obtained from  \eq{centradeutp}, \eq{regdetpot} by the substitution $g_{\pi NN}^2/M_N^2\to g_A^{H_1}
g_A^{H_2}/F_\pi^2$, where the constants $g_A^{H_i}$ are axial charges of the respective hadrons.
The axial charges of heavy charmed hadrons almost never can be derived from the experimental data. An
estimate of these charges can be obtained with the help of the naive constituent quark model for
light quarks as suggested in \cite{dorgeb2001}. We do not expect these axial charges to be
particularly accurate, but they would have at least correct signs and order of magnitude. In the
leading nonrelativistic approximation the time component of the quark axial current $j^5_{\mu
a}=\bar\psi_qt_a\gamma_\mu\gamma^5\psi$ turns into zero, and only the spatial components of the
axial current proportional to $\sigma_i$ survive. Then in the framework of the nonrelativistic
constituent quark model the axial charge $g_{A}^H$ of the hadron $H$ is given by the relationship

\beq \label{axialcharg} g_{A}^HS_{i}T_{a}=g_{A}^q\left\langle\Psi_{H}\left|\sum
s_{i}t_{a}\right|\Psi_{H}\right\rangle, \eeq

\noindent where $|\Psi_{H}\rangle$ is the heavy hadron state vector in terms of the light quark
states, $g_A^q$ is the light quark axial charge, $s_i$ and $t_a$ are the quark spin
and isospin operators, and $S_i$ and $T_a$ are the respective heavy hadron
operators. Summation goes over all light constituent quarks. It is easy to generalize this
expression for transitional axial charges $g_{A}^{H_1H_2}$.

Below we will assume that $g_A^q=1$ (and $g_A^{\bar q}=-1$ for antiquarks). We can make a more
accurate estimate of the quark axial charge using the Goldberger-Treiman relationship and
\eq{axialcharg} for the nucleon axial charge. In this way we would obtain

\beq g_A^q\approx\frac{3}{5}\frac{g_{\pi NN}F_\pi}{M_N}=0.81. \eeq

\noindent Taking into account inaccuracy of the nonrelativistic constituent quark model itself it
does not make much sense to make a distinction between $g_A^q=1$ and $g_A^q=0.81$. Axial charges of
the nucleon and some charmed hadrons are collected  in Table~\ref{axchrg}.

\begin{table}[h!]
\caption{\label{axchrg} Some axial charges}
\begin{ruledtabular}
\begin{tabular}{lc}
$H$ & $g_A^H$
\\
\hline $N$& $\frac{5}{3}$
\\
$\bar D$&  $\frac{1}{2}$
\\
$\bar D^*$&  $\frac{1}{2}$
\\
$\Sigma_c$ &  $\frac{2}{3}$
\\
$\Lambda_c$ &  $0$
\\
$\Sigma_c^*$ &  $\frac{1}{3}$ 
\end{tabular}
\end{ruledtabular}
\end{table}

Let us return to the one-pion exchange interaction between $\Sigma_c$ and $\bar D^*$.  Unlike the
case of the deuteron the tensor potential in this case does not commute with the  total spin $\bm
S=\bm S_{\Sigma_c}+\bm S_{\bar D^*}$, $[S_{12}(\bm S_{\Sigma_c},\bm S_{\bar D^*},\bm n),\bm S]\neq
0$. Only  the total angular momentum $\bm J=\bm L+\bm S$ is conserved. The lowest $\Sigma_c\bar D^*$
bound state should be dominated by the $S$-wave, and seeking an interpretation for the $P_c(4450)$
pentaquark we start with the sector with $J=3/2$. The tensor potential has nonzero matrix elements
between  $S$- and $D$-waves and the $\Sigma_c\bar D^*$ state with $J=3/2$ is a superposition of
three states $|L=0,S=3/2\rangle$, $|L=2,S=1/2\rangle$, and $|L=2,S=3/2\rangle$. The Hamiltonian  in
the subspace with $J=3/2$ and $T=1/2$ has the form (it coincides with the pion contribution to the
respective Hamiltonian in \cite{yshlz2012})

\beq \label{pcpionham} H=\left(
\begin{array}{c|c|c}
-\frac{\nabla^2}{2\mu}+V_C&-\frac{1}{2}V_T&V_T\\ \hline
-\frac{1}{2}V_T&-\frac{\nabla^2}{2\mu}+\frac{3}{\mu r^2}-2V_C&\frac{1}{2}V_T\\ \hline
V_T&\frac{1}{2}V_T&-\frac{\nabla^2}{2\mu}+\frac{3}{\mu r^2}+V_C
\end{array}
\right), \eeq

\noindent where $\mu$ is the $\Sigma_c\bar D^*$ reduced mass.   The potentials  (nonregularized and
regularized, with the subtracted $\delta$-function contribution) according to \eq{centradeutp} and
\eq{regdetpot} are

\beq \label{regnregpot}
\begin{split}
V_{C}(r)=&-\frac{m_{\pi}^{2}}{F_{\pi}^{2}}\frac{e^{-m_{\pi}r}}{18\pi r},
\qquad\qquad\qquad\qquad\qquad
V_{C,reg}(r)=-\frac{m_{\pi}^{2}}{F_{\pi}^{2}}\frac{1}{18\pi}Y(\Lambda,m_\pi,r),\\
V_{T}(r)=&-\frac{1}{F_{\pi}^{2}}(3+3m_\pi r+m_\pi^{2}r^{2})\frac{e^{-m_{\pi}r}}{9\pi r^{3}},\qquad
V_{T,reg}(r)=-\frac{1}{F_{\pi}^{2}}\frac{1}{9\pi}Z(\Lambda,m_\pi,r).
\end{split}
\eeq

\noindent In this calculation we used $g^{\Sigma_c}_A=2/3$, $g^{\bar D^*}_A=1/2$ from
Table~\ref{axchrg} and the substitution $g_{\pi NN}^2/M_N^2\to g_A^{H_1} g_A^{H_2}/F_\pi^2$
discussed above.

The spin-spin $S$-wave potential is attractive and is suppressed  by the factor $8/25$ in comparison
with the respective deuteron potential. Again, like in the deuteron case it is not strong enough to
bind  $\Sigma_c$ and $\bar D^*$ if the tensor potential is turned off.

\begin{figure}[h!]
\center\epsfig{file=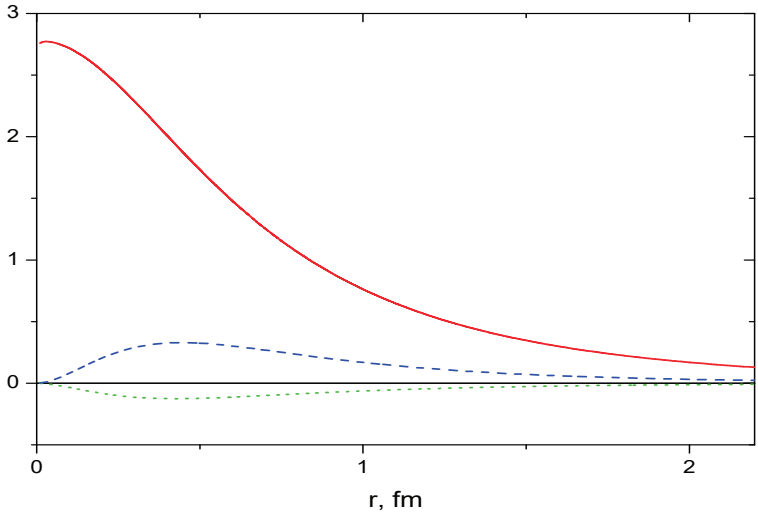
,height=8cm, width=14cm
}
\caption{\label{normwfpent}Normalized wave
functions of the $\Sigma_c\bar D^*$ bound state with  $J^P=3/2^-$,  $T=1/2$ and the binding energy
14.6 MeV ($\Lambda=1430$ MeV). The $|L=0,S=3/2\rangle$, $|L=2,S=1/2\rangle$, and $|L=2,S=3/2\rangle$ wave functions are solid, dotted, and dashed lines, respectively. }
\end{figure}

We looked for the bound state solutions with the Hamiltonian in \eq{pcpionham}. To cut off the
singular behavior of the potential at short distances we one time amended the potential in
\eq{pcpionham} by an infinitely hard wall at small distances, and another time we used the
regularized potential in \eq{regnregpot}. In a model with the wall at $r_0=0.33$ fm we find a
$\Sigma_c\bar D^*$  bound state with  $J^P=3/2^-$,  $T=1/2$ and the binding energy 14.7 MeV, exactly with the mass of the $P_c(4450)$  pentaquark. This is a deuteronlike state, the binding arises due
to the nondiagonal tensor potential.  The hard core radius $r_0=0.33$ fm is somewhat smaller than in the case of the deuteron, but is still not too small.  The rms of the bound state is about 1.6 fm.
This radius is large in comparison with the hard core radius and with the scale corresponding to the exchanges by other light bosons, what justifies the one-pion exchange approximation. The fraction of the $D$-wave squared is about 18\%, much larger than in the deuteron. In a model with the
regularized potentials \eq{regnregpot} the bound state at the position of $P_c(4450)$ arises at
$\Lambda=1430$ MeV. The rms in this case is about 1.24 fm and the fraction of $D$-wave squared is
12\%, see wave functions in Fig.~\ref{normwfpent}. These results were obtained with the axial
charges in Table~\ref{axchrg}. We also repeated these calculations with the phenomenological axial
charges, see, e.g., \cite{yshlz2012}. We again can obtain a bound state at the position of
$P_c(4450)$ but now $\Lambda=2000$ MeV, rms is 1.13 fm, and the fraction of the $D$-wave squared is
about 10\%. Dependence of the energy level on $\Lambda$ with any choice of the set of coupling
constants is rather steep, the binding energy changes by about 4 MeV when $\Lambda$ changes by 100
MeV.  We see that $P_c(4450)$ can be interpreted as a deuteronlike bound state, but requires fine
tuning of the short distance regularization parameter $\Lambda$.

Now, that we fixed $\Lambda$, it is natural to look for the $\Sigma_c\bar D^*$ bound states in the
channels with $J=3/2$, $T=3/2$ and $J=1/2$, $T=1/2,3/2$. It turns out that there are no
$\Sigma_c\bar D^*$  bound states with other quantum numbers besides $J^P=3/2^-$,  $T=1/2$ for the
values of $\Lambda$ and/or the position of the hard wall determined above.

A $\Sigma_c\bar D^*$ bound state with $J^P=3/2^-$,  $T=1/2$  and the binding energy 85 MeV was
obtained in \cite{cllz2015,clz2016} on the basis of the one-pion exchange. It was identified with
the $P_c(4380)$ LHCb pentaquark. The binding  in \cite{cllz2015,clz2016} occurred due to the spin-spin part of the one-pion exchange potential with account only for the $S$-wave wave function. The one-pion potential in \cite{cllz2015} was regularized  by the dipole form factor with $\Lambda=2.35$ GeV  ($\Lambda=1.78$
GeV in \cite{clz2016}). The sign of the unregularized spin-spin potential in \cite{cllz2015,clz2016}
is opposite to the sign in \eq{pcpionham} and in \cite{yshlz2012}, and corresponds to a long-distance
repulsion. The binding is due to the regularized $\delta$-function contribution to the spin-spin
potential, see \eq{centradeutp} (compare with the regularized potential in \eq{regnregpot}, where
the $\delta$-function contribution is subtracted). The repulsive long-distance one-pion spin-spin
contribution to the potential in \cite{cllz2015,clz2016} can be omitted without changing the
results. The rms of the bound state in \cite{cllz2015,clz2016} is 0.42 fm, what is by far too small
to justify validity of the one-pion exchange approximation\footnote{We reproduced the calculations
in \cite{cllz2015,clz2016} and discovered some misprints. In particular, the horizontal axis in
Fig.1 in \cite{cllz2015} and in Fig.3 in \cite{clz2016} is graduated in $GeV^{-1}$ not in fm.}. This
rms value can be roughly estimated almost without calculations, simply from the well known
asymptotic formula for the bound state wave function $\psi(r)_{r\to\infty}\sim
e^{-\frac{r}{r_0}}\sim e^{-r\sqrt{2\mu E}}$, where $E$ is the binding energy. We disagree with the
sign of the one-pion potential in \cite{cllz2015,clz2016}, do not accept the idea of binding due to
the attractive smeared $\delta$-function, cannot justify the dominant role of the one-pion exchange
at the distances about 0.4 fm, and therefore cannot accept the interpretation of  $P_c(4380)$ as a
$\Sigma_c\bar D^*$ bound state due to the one-pion exchange.

\section{Are there other deuteronlike pentaquarks?}

In search for other loosely bound pentaquark states with a deuteronlike binding we considered
possible $\Sigma_{c}^{\ast}\bar{D}^{\ast}$ loosely bound states. One could expect to find bound
states with the binding energy about 10 MeV and mass about 4520 MeV, slightly below
$M_{\Sigma_{c}^{\ast}}+M_{\bar{D}^{\ast}}=4530$ MeV.

Like in the case of  $\Sigma_c\bar D^*$ interaction only the total angular momentum is conserved. We
consider first the sector with $J=5/2$.  The lowest $\Sigma^*_c\bar D^*$ bound state should be
dominated by the $S$-wave, and the tensor potential has nonzero matrix elements between  $S$- and
$D$-waves. Hence, the $\Sigma^*_c\bar D^*$ state with $J=5/2^-$ is a superposition of four partial
waves, $|L=0,S=5/2\rangle$, $|L=2,S=1/2\rangle$, $|L=2,S=3/2\rangle$, and $|L=2,S=5/2\rangle$. The
Hamiltonian in the subspace with $J=5/2$, $T=1/2$ has the form

\beq \label{pcpionhamheav} H=\left(
\begin{array}{c|c|c|c}
-\frac{\nabla^2}{2\mu}+V_C&-\sqrt{\frac{3}{5}}V_T&\frac{\sqrt{21}}{10}V_T&\frac{3}{5}\sqrt{14}V_T\\
\hline -\sqrt{\frac{3}{5}}V_T&-\frac{\nabla^2}{2\mu}+\frac{3}{\mu
r^2}-\frac{5}{3}V_C&\frac{1}{2}\sqrt{\frac{7}{5}}V_T&2\sqrt{\frac{6}{35}}V_T\\ \hline
\frac{\sqrt{21}}{10}V_T&\frac{1}{2}\sqrt{\frac{7}{5}}V_T &-\frac{\nabla^2}{2\mu}+\frac{3}{\mu
r^2}-\frac{2}{3}V_C+\frac{8}{7}V_T&-\frac{1}{7}\sqrt{\frac{3}{2}}V_T\\ \hline
\frac{3}{5}\sqrt{14}V_T&2\sqrt{\frac{6}{35}}V_T&-\frac{1}{7}\sqrt{\frac{3}{2}}V_T&-\frac{\nabla^2}{2\mu}
+\frac{3}{\mu r^2}+V_C+\frac{6}{7}V_T
\end{array}
\right). \eeq

\noindent The potentials (nonregularized and regularized, with the subtracted $\delta$-function
contribution) according to \eq{centradeutp} and \eq{regdetpot} are

\beq \label{regnregpothaev}
\begin{split}
V_C(r)=&-\frac{m_{\pi}^{2}}{F_{\pi}^{2}}\frac{e^{-m_{\pi}r}}{12\pi r},
\qquad\qquad\qquad\qquad\qquad
V_{C,reg}(r)=-\frac{m_{\pi}^{2}}{F_{\pi}^{2}}\frac{1}{12\pi}Y(\Lambda,m_\pi,r),\\
V_{T}(r)=&-\frac{1}{F_{\pi}^{2}}(3+3m_\pi r+m_\pi^{2}r^{2})\frac{e^{-m_{\pi}r} }{18\pi r^{3}},
\qquad V_{T,reg}(r)=-\frac{1}{F_{\pi}^{2}}\frac{1}{18\pi}Z(\Lambda,m_\pi,r).
\end{split}
\eeq

\noindent In this calculation we used $g^{\Sigma_c^*}_A=1/3$, $g^{\bar D^*}_A=1/2$ from
Table~\ref{axchrg} and the substitution $g_{\pi NN}^2/M_N^2\to g_A^{H_1} g_A^{H_2}/F_\pi^2$
discussed above.

\begin{figure}[h!]
\center\epsfig{file=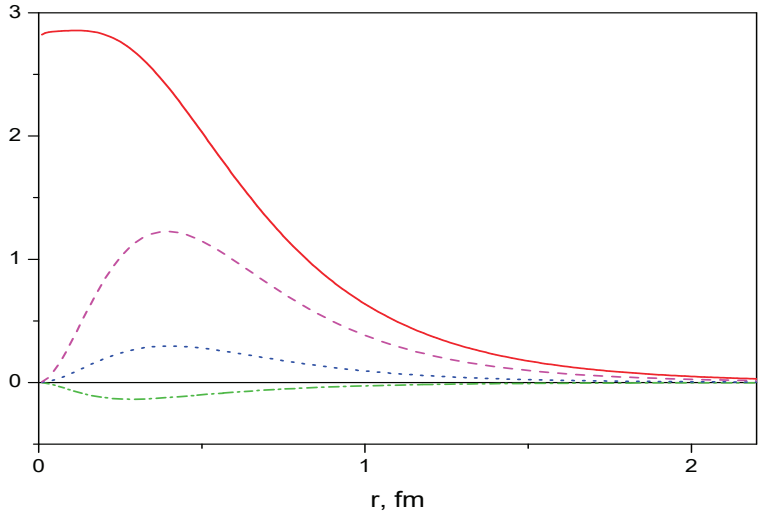
,height=8cm,width=14cm
}
\caption{\label{normwfpenth}Normalized wave
functions of the $\Sigma_c^*\bar D^*$ bound state with  $J^P=5/2^-$,  $T=1/2$ and the binding energy
74.8 MeV ($\Lambda=1400$
MeV). The $|L=0,S=5/2\rangle$, $|L=2,S=1/2\rangle$, $|L=2,S=3/2\rangle$, and $|L=2,S=5/2\rangle$
wave functions are solid, dash-dotted, dotted, and dashed lines, respectively. }
\end{figure}

The spin-spin $S$-wave potential is attractive and is $1.5$ times stronger than the respective
potential in the case of $\Sigma_c\bar D^*$. We looked for the eigenstates of the Hamiltonian in
\eq{pcpionhamheav} with the hard wall at $r_0=0.33$ fm, exactly at the same position as in the case
of $\Sigma_c\bar D^*$ above. Surprisingly, there are no shallow loosely bound states with the
binding energy about dozens of MeV. Only a bound state with the binding energy $82$ MeV and the mass
close to the mass of $P_c(4450)$ exists. There is a steep dependence of the binding energy on the
radius of the hard core. Change of this radius from $r_0=0.33$ fm to $r_0=0.38$ fm reduces the
binding energy to 10 MeV and leads to prediction of a pentaquark state with $J^P=5/2^-$, $T=1/2$ and
mass 4520 MeV. We repeated these calculations with the phenomenological coupling constants (see,
e.g., \cite{yshlz2012}) and the regularized potentials in \eq{regnregpothaev}. Again we observe a
steep dependence of the binding energy on the regularization parameter  $\Lambda$, see
Table~\ref{ebfroml}. At $\Lambda=1400$ we obtain a state with the mass of $P_c(4450)$, $J^P=5/2^-$
and $T=1/2$. Recall that we have already found a $\Sigma_c\bar D^*$ bound state with the mass of
$P_c(4450)$, $J^P=3/2^-$ and $T=1/2$ at $\Lambda=2000$ MeV using the phenomenological coupling
constants. Interpretation of the $\Sigma^*_c\bar D^*$ bound state as $P_c(4450)$ could be
preferable, since $\Lambda$ is smaller than in the case of $\Sigma_c\bar D^*$. On the other hand rms
of the  $\Sigma^*_c\bar D^*$ bound state at $\Lambda=1400$ MeV is only 0.78 fm, and the fraction of
the $D$-wave squared is about 25\%, see Fig.~\ref{normwfpenth}. The one-pion exchange mechanism is
probably not dominant for the bound state with such rms. The steep dependence of the bound state
mass on the magnitude of $\Lambda$  makes prediction of the mass of this state not too reliable.  A
fair conclusion could be that the consideration above does not have too much predictive power.

\begin{table}[h!]
\caption{\label{ebfroml} Binding energy of $\Sigma_c^*\bar D^*$ ($J^p=/5/2^-$, $T=1/2$) state on
$\Lambda$}
\begin{ruledtabular}
\begin{tabular}{ll}
$\Lambda$ (MeV)& $E_b$ (MeV)
\\
\hline $800$& $-1.18$
\\
\hline $900$&  $-4.11$
\\
$1000$&  $-9.64$
\\
$1100$ &  $-18.6$
\\
$1200$ &  $-31.8$
\\
$1300$ &  $-50.3$
\\
$1400$&$-74.8$
\end{tabular}
\end{ruledtabular}
\end{table}

We also looked for possible $\Sigma^*_c\bar D^*$ bound states in the channels with $J=3/2$, $T=3/2$,
and $J=1/2$, $T=1/2,3/2$. We found a shallow bound state with $J=1/2$, $T=3/2$, binding energy
$E_b=-1.4$ MeV, mass 4526 MeV,  rms about 3 fm, fraction of the $D$-wave squared about 6\%  at
$\Lambda=1400$ MeV. This is an almost ideal deuteronlike bound state, but again a steep dependence
on $\Lambda$ does not allow to make a reliable prediction.

The $\Sigma_c^*\bar D^*$ one-pion exchange interaction was also considered in
\cite{cllz2015,clz2016}, where a bound state with the binding energy 70 MeV, $J^P=5/2^-$, $T=1/2$,
and mass 4450 MeV was obtained and identified with $P_c(4450)$. Like in the case of $\Sigma_c\bar
D^*$ we disagree with the sign of the potential in \cite{cllz2015,clz2016}, see discussion above.

\section{Nuclear  Type Potential and  Pentaquarks}

We considered above the deuteronlike model of loosely bound pentaquarks where the long distance
one-pion exchange amended by a short distance repulsive core  plays the defining role. We have
already mentioned that more sophisticated descriptions of nucleon-nucleon interactions include
exchanges by other light mesons, see, e.g., \cite{bertul2007}. We do not need to go into detail of
these more sophisticated models, but let us try to include exchanges by $\sigma$, $\rho$, $\omega$,
and $\eta$ in the most straightforward way. One can use the constituent quark model to estimate the
meson-nucleon interaction constants, see, e.g., \cite{dly2009}. In the framework of this model,
$\sigma$ and $\omega$ interact with the light quark baryon number and $\rho$ interacts with isospin.
Exchange by a scalar $\sigma$ is always attraction, by a vector $\omega$ it is always repulsion and
by a vector $\rho$ it depends on isospin. In the case of the deuteron the total effect of the $\rho$
and $\omega$ exchanges is repulsion, that naturally explains the origin of the repulsive core at
small distances. The regularization at small distances is still needed due to the singular behavior
of the tensor potential at small distances. The deuteron problem with the nuclear type potential and
phenomenological coupling constants was considered in \cite{dly2009}. It turned out that inclusion
of the meson exchanges effectively did not change values of the deuteron parameters obtained in the
one-pion exchange scenario and the value of the regularization parameter $\Lambda$ also did not
change. This justifies the one-pion binding scenario for the deuteron.

Almost all papers on the molecular model of tetra- and pentaquarks adopt the nuclear point of view
and construct nuclear type potentials that include exchanges by all light bosons, see, e.g., review
in \cite{cclz2016}. Like in nuclear physics each meson exchange is regularized at short distances by
the dipole potential with a phenomenological  cutoff parameter $\Lambda$. This potential is
nonuniversal and the cutoff parameter is adjusted for each experimentally known tetra- or
pentaquark.

Like in the case of the deuteron one way to test the reliability of the one-pion exchange scenario
for pentaquarks considered above is to include in the potential exchanges by other mesons and to see
if the characteristics of the bound states (rms radius, fraction of the $D$-wave squared, parameter
$\Lambda$)  would remain stable. We obtained above a $\Sigma_c\bar D^*$ ($J^P=3/2^-$, $T=1$) bound
state at $\Lambda=2000$ MeV with mass $4450$ MeV, rms 1.13 fm, the fraction of the $D$-wave squared
about 10\%. Now we amend the one-pion exchange potential by $\sigma$, $\rho$, $\omega$, and $\eta$
exchanges, and use the phenomenological coupling constants (but not the signs of the individual
contributions to the potential and not the regularization without subtraction of the
$\delta$-function contribution) from \cite{yshlz2012}. Then the bound state at the position of
$P_c(4450)$ arises at $\Lambda=1300$ MeV with rms 1.46 fm, and $D$-wave squared fraction about 4\%.
The value of $\Lambda$ changed significantly thus hinting that the one-pion mechanism is not too
reliable for this bound state.

We have also reconsidered $\Sigma^*_c\bar D^*$ bound states in a model with the nuclear type
potential. Like in the one-pion scenario we obtain a $J^P=5/2^-$, $T=1/2$ bound state with
$E_b=-74$ MeV, rms 0.71 fm, and $D$-wave squared fraction 15\% at $\Lambda=1200$ MeV.  Again, like
in the one-pion exchange scenario we have also found a second $\Sigma^*_c\bar D^*$ bound state at
the same $\Lambda$. This is $J^P=1/2^-$, $T=3/2$  state with $E_b=-37$ MeV, rms 0.83 fm, and
$D$-wave squared fraction 1.2\%. We observe a rather substantial change of parameters, especially of
the second state,  in comparison with the one-pion exchange scenario.

We see that the one-pion exchange scenario is not stable with respect to inclusion of other light
meson exchanges, and it is hard to insist that this is the dominant binding mechanism even for the
loosely bound pentaquarks. Both the one-pion exchange and nuclear type scenarios for pentaquarks
suffer from steep dependence on the short distance regularization parameter $\Lambda$  (or position
of the hard wall at small distances). One can describe existing experimental data on pentaquarks with the help of nuclear type and/or one-pion exchange potentials choosing different values of $\Lambda$ for different states.

There are also apparent problems with the pentaquark decays in the molecular approach. The
pentaquarks were discovered in the invariant mass spectrum of $J/\psi$ and the proton. So decay to
$J/\psi+N $ is the only reliably established pentaquark decay mode.  It is hard to understand how
this decay to the states without open charm can can give a substantial contribution to the total
width in the molecular picture.  The constituents with open charm preserve their individuality in
the molecular picture and we expect  the decays into states without open charm to be strongly
suppressed.  This presents a qualitative difference with the hadrocharmonium picture, where the
charmed quarks are close to each other and decay into $J/\psi+N$ should  should give a significant
contribution.

\section{Discussion of Results}

We have developed a QCD based approach to dynamical interpretation of pentaquarks. In this approach
pentaquarks arise as bound states of ordinary baryons and excited states of quarkonia. The binding
is due to the chromoelectric dipole interaction between the quarkonia states and ordinary baryons.
The strength of the quarkonium-baryon interaction is determined by the quarkonium state
chromoelectric polarizability and the baryon mass. The interaction potential is proportional to the
density of the baryon energy-momentum distributions. In this approach we interpret the LHCb
$P_c(4450)$ pentaquark as a bound $\psi'N$ state with spin-parity $J^P=3/2^-$. We calculated its
decay width into $J/\psi N$, $\Gamma(P_c(4450)\to J/\psi+N)\approx 11$ MeV, what is in rough
agreement with the experimental data. The $P_c(4450)$ pentaquark in this approach turns out to be a
member of  a pentaquark flavor octet, similar to the octet of ordinary baryons. The interaction
between the quarkonia states and ordinary baryons is spin-independent in the leading approximation,
so there are two degenerate pentaquark octets with spin-parities $J^P=1/2^-$ and $J^P=3/2^-$. This
degeneracy is lifted by a small color-singlet spin-spin interaction due to the interference of the
chromoelectric $E1$ and chromomagnetic $M2$ transitions in charmonium.

Experimentally  acceptable spin-parities for the LHCb pentaquarks are $(3/2^-,5/2^+)$,
$(3/2^+,5/2^-)$, and $(5/2^+,3/2^-)$ \cite{LHCb2015}. With our assignment of spin-parity $J^P=3/2^-$
to $P_c(4450)$ we have to assign $J^P=5/2^+$ to the wide $P_c(4380)$. We cannot find a natural
interpretation for such state as a $\psi'N$ bound state. We have discovered $\chi_{c0}N$,
$\chi_{c1}N$, $\chi_{c2}N$, and $h_c N$ bound states with spin-parity $J^P=5/2^+$ in the same mass
region as $P_c(4380)$. Unfortunately, these states have no open channels for decays into an ordinary
baryon and a charmonium state with hidden charm. Hence, all strong decays should go via heavy
quark-antiquark annihilation, and are thus strongly suppressed in accordance with the
Okubo-Zweig-Iizuka rule.

From the phenomenological perspective the main problem with the hadrocharmonium approach is its
apparent inability to describe the $P_c(4380)$ pentaquark. On the theoretical side, development of a
reliable method to calculate charmonium polarizability is urgently required.

We also considered the one-pion exchange model for loosely bound pentaquarks. In the framework of
this model the $P_c(4450)$ pentaquark could be interpreted as a $\Sigma_c\bar D^*$ bound state with
$J^P=3/2^-$ and $T=1/2$. The $P_c(4380)$ pentaquark that did not find a satisfactory explanation in
the hadrocharmonium approach, does not admit a description based on the one-pion exchange either.
Another difficulty  of the one-pion exchange model is connected with decays. Phenomenologically we
should expect that the partial width $\Gamma(P_c(4450)\to J/\psi+ N)$ is not small. This follows
from the simple observation that the $P_c(4450)$ pentaquark was discovered in the invariant mass
distribution of $J/\psi N$. The amplitude of this decay is proportional to the bound state wave
function at zero, since to create $J/\psi$ the charmed quark and antiquark should come closely
together. On the other hand the pentaquark in the one-pion molecular model is loosely bound, the
wave function is smeared over a relatively large region and as a result the bound state wave
function at zero is relatively small. Suppression of the pentaquark $J/\psi+ N$ decay mode is an
apparent difficulty of the one-pion exchange model.

A nuclear type potential, that includes exchanges of different light mesons allows to consider not
necessarily loosely bound pentaquarks. In this approach the binding energy could be relatively large
and the constituents could be at much shorter distances than in the  one-pion exchange model. This
approach is much more flexible than the deuteronlike mechanism. The problem of the nuclear type
potential which it shares with the one-pion potential is the apparent lack of predictive power.  One
cannot describe even the already observed pentaquarks with one and the same  short distance
regularization. An adjustment of this parameter  is required for each particular state.

Either of the scenarios above predicts a number of new pentaquark states. A hadrocharmonium   LHCb pentaquark is a member  of an $SU(3)$ flavor octet with small mass splittings (see the discussion above). The situation with a molecular type pentaquark is less clear. The $SU(3)$ partners of such pentaquark can fail to form a bound state. This  is what happens in the case of the deuteron, where partners do not form due to the difference of the pion and kaon masses. The pentaquark binding energy is larger, constituents in the pentaquarks are more densely packed than in the deuteron, so the mass differences inside the meson octets are less important. Whether the molecular pentaquarks have the $SU(3)$ partners remains an open question, that probably cannot be answered in our rough approximation.

It is interesting to explore if there exist molecular bound states of other heavy mesons and baryons due to the one-pion exchange mechanism. As a simplest possibility it was suggested some time ago \cite{hllz2006} that $\Lambda_c(2940)$ could  be a molecular pentaquark made of $\bar{D}^*$ and nucleon. We tested this suggestion quantitatively in the same approach as in the discussion of the $P_c(4450)$ pentaquark. It turned out that there is no bound state in the $N\bar D^*$ channel. This happens because due to the isospin factors the attraction induced by the pions in the $N\bar D^*$ channel  is only $3/4$ of the attraction  between $\Sigma$ and $\bar{D}^*$, while the nucleon is lighter than $\Sigma_c$ (the coupling constants in both channels are approximately equal). Still, molecular pentaquarks formed by heavier baryons with different quantum numbers and by other heavy mesons could exist.

Hadrocharmonium scenario also admits existence of pentaquarks formed by other baryons (for example, by the Roper resonance as suggested in \cite{kv2016}) and  different $c\bar c$ states. The chromoelectric interaction is spin-independent, so the bound states in this case (if any) will come in multiplets splitted by hyperfine interaction.  They also should form the $SU(3)$ flavor multiplets.

Both the hadrocharmonium interpretation of pentaquarks and the molecular-like approach have their
own drawbacks and advantages, and need further development. Experimental and theoretical research on pentaquark decay rates and branching ratios could help to discriminate between different models. We
hope to address decays  in the future.

\acknowledgments

This paper was supported by the NSF grants PHY-1402593 and PHY-1724638.  The work of M.~V.~P. is supported by CRC110 of DFG.


\begin{thebibliography}{99}

\bibitem{jmr2016} J.-M.~Richard, Few Body Syst. \textbf{57}, 1185 (2016).

\bibitem{LHCb2015} R.~Aaij {\it et al.} [LHCb Collaboration], Phys. Rev.  Lett.  \textbf{ 115},
    072001 (2015).

\bibitem{dubvol2008} S.~Dubynskiy, M.~B.~Voloshin, Phys. Lett. B \textbf{ 666}, 344 (2008)

\bibitem{sibvol2005} A.~Sibirtsev and M.~B.~Voloshin, Phys. Rev. D \textbf{71}, 076005 (2005).

\bibitem{livol2014} X.~Li, M.~B.~Voloshin, Mod. Phys. Lett. A \textbf{29}, 1450060 (2014).

\bibitem{epp2016}  M.~I.~Eides, V.~Yu.~Petrov, and M.~V.~Polyakov, Phys. Rev. D \textbf{93}, 054039
    (2016).

\bibitem{PPS} I.~A.~Perevalova, M.~V.~Polyakov, and P.~Schweitzer, Phys. Rev. D \textbf{94}, 054024
    (2016).

\bibitem{sjbisgt} S.~J.~Brodsky, I.~Schmidt, and G.~F.~de~Teramond, Phys. Rev. Lett. \textbf{64}, 1011 (1990).

\bibitem{lms1992} M.~Luke, A.~V.~Manohar, and M.~J.~Savage, Phys. Lett. B \textbf{288}, 355 (1992).

\bibitem{volsh2008}  M.~B.~Voloshin, Prog. Part. Nucl. Phys. \textbf{61}, 455 (2008).

\bibitem{volok1976} M.~B.~Voloshin and L.~B.~Okun, Pis'ma Zh. Eksp. Teor. Fiz. 23, 369 (1976) [JETP
    Lett. 23, 333 (1976)].

\bibitem{cclz2016} H.-X.~Chen, W.~Chen, X.~Liu, S.-L.~Zhu, Phys.~Rep. \textbf{639}, 1 (2016).

\bibitem{torn1991} N.~A.~T\"ornqvist, Phys. Rev. Lett. \textbf{67}, 556 (1991).

\bibitem{teoegk1993} T.~E.~O.~Ericson and G.~Karl, Phys. Lett. B \textbf{309}, 426 (1993).

\bibitem{torn1994} N.~A.~T\"ornqvist, Z. Phys. C \textbf{61}, 525 (1994).

\bibitem{bertul2007} C.~A.~Bertulani, {\it Nuclear Physics in a Nutshell}, (Princeton University
    Press, Princeton and Oxford, 2007).

\bibitem{mpr2015} L.~Maiani, A.~D.~Polosa and V.~Riquer, Phys. Lett. B \textbf{ 749},  289 (2015).

\bibitem{amnss2015} V.~V.~Anisovich, M.~A.~Matveev, J.~Nyiri, A.~V.~Sarantsev, and A.~N.~Semenova,
    arXiv:1507.07652.

\bibitem{leb2015} R.~F.~Lebed,  Phys. Lett. B \textbf{749},  454 (2015),

\bibitem{lhh2015} G.~N.~Li, M.~He and X.~G.~He, JHEP \textbf{12}, 128 (2015).

\bibitem{hycckc2015} H.-Y.~Cheng and C.-K. Chua, Phys. Rev. D \textbf{92}, 096009 (2015).

\bibitem{aaar2016} A.~Ali, I.~Ahmed, M.~J.~Aslam, and A.~Rehman, Phys. Rev. D \textbf{94}, 054001 (2016).

\bibitem{als2017}  A.~Ali, J.~S.~Lange, and S.~Stone, Prog.Part.Nucl.Phys. \textbf{97}, 123 (2017).

\bibitem{amam2015} A.~Mironov and A.~Morozov, JETP Lett. \textbf{102}, 271 (2015).

\bibitem{ugmjao2015}  U.~G.~Mei\ss ner and J.~A.~Oller, Phys. Lett. B \textbf{751}, 59 (2015).

\bibitem{mikh2015}  M.~Mikhasenko, arXiv:1507.06552.

\bibitem{gmwy2015} F.~K.~Guo, U.~G.~Mei\ss ner, W.~Wang, and Z.~Yang, Phys. Rev. D \textbf{92},
    071502 (2015).

\bibitem{lreo2016}  L.~Roca and E.~Oset, Eur. Phys. J. C \textbf{76}, 591 (2016).

\bibitem{kgottf1978} K.~Gottfried, Phys. Rev. Lett. \textbf{40}, 598 (1978).

\bibitem{mbv1979} M.~B.~Voloshin, Nucl. Phys. B \textbf{154} 365, 1979.

\bibitem{pesk1979} M.~Peskin,  Nucl. Phys. B \textbf{156}, 365 (1979).

\bibitem{bp1979} G.~Bhanot and M.~E.~Peskin, Nucl. Phys. B \textbf{156}, 391 (1979).

\bibitem{leut1981} H.~Leutwyler, Phys. Lett. B \textbf{98},  447 (1981).

\bibitem{volosh1982} M.~B.~Voloshin, Sov. J. Nucl. Phys. \textbf{36}, 143  (1982).

\bibitem{bramb2015} N.~Brambilla, G.~Krein, J.~Castell\'a, and A.~Vairo, Phys. Rev. D \textbf{93},
    054002 (2016).

\bibitem{volosh2004} M.~B.~Voloshin, Mod. Phys. Lett. A \textbf{19}, 665 (2004).

\bibitem{ns1981}  V.~A.~Novikov and M.~A.~Shifman, Z. Phys. C \textbf{ 8}, 43 (1981).

\bibitem{maxim2007} K.~Goeke, J.~Grabis, J.~Ossmann, M.~V.~Polyakov, P.~Schweitzer, A.~Silva, and
    D.~Urbano, Phys. Rev. D \textbf{75}, 094021 (2007).

\bibitem{tc2008} C.~E.~Thomas and F.~E.~Close, Phys. Rev. D \textbf{78}, 034007 (2008).

\bibitem{lzah2016} Y. Liu and I. Zahed, Phys. Lett. B \textbf{762}, 362 (2016).

\bibitem{skcb2003} S.~K.~Choi {\it et al.} [Belle Collaboration],  Phys. Rev. Lett. \textbf{91},
    262001 (2003).

\bibitem{torn2004} N.~A.~T\"ornqvist, Phys. Lett. \textbf{590}, 209 (1991).

\bibitem{ssh2016} Y.~Shimizu, D.~Suenaga, and M.~Harada,  Phys. Rev. D \textbf{93}, 114003 (2016).

\bibitem{dorgeb2001} D.~O.~Riska  and  G.~E.~Brown, Nucl. Phys. A \textbf{679}, 577 (2001).

\bibitem{yshlz2012}  Z.-C.~Yang, Z.-F.~Sun, J.~He, X.~Liu, and S.-L.~Zhu, Chin. Phys. C \textbf{36},
    6 (2012).

\bibitem{cllz2015} R.~Chen, X.~Liu, X.-Q.~Li, and S.-L.~Zhu, Phys. Rev. Lett. \textbf{115}, 132002
    (2015).

\bibitem{clz2016} R.~Chen, X.~Liu, and S.-L.~Zhu, Nucl. Phys. A \textbf{954}, 406 (2016).

\bibitem{dly2009} G.-J.~Ding, J.-F.~Liu, and M.-L.~Yan,  Phys. Rev. D \textbf{79}, 054005 (2009).

\bibitem{hllz2006} X.-G.~He, X.-Q.~Li, X.~Liu, and X.-Q.~Zeng, Eur. Phys. J. C \textbf{51}, 883  (2007).

\bibitem{kv2016} V.~Kubarovsky and  M.~B.~Voloshin, Phys. Rev. D \textbf{92}, 031502 (2015).


\end{thebibliography}
\end{document}